# Selecting Differential Splicing Methods: Practical Considerations

Ben J. Draper*[1], Mark J. Dunning[2], David C. James[1]

[1] Department of Chemical and Biological Engineering, University of Sheffield, Mappin St., Sheffield, S1 3JD, U.K.

[2] Bioinformatics Core, The Faculty of Health, The University of Sheffield, Sheffield, S10 2HQ, U.K.

*To whom correspondence should be addressed.

telephone: +44 (0)7305 179653

email: bjdraper1@sheffield.ac.uk

ORCID ID: 0009-0005-8034-0172

## Abstract

Alternative splicing is crucial in gene regulation, with significant implications in clinical settings and biotechnology. This review article compiles bioinformatics RNA-seq tools for investigating differential splicing; offering a detailed examination of their statistical methods, case applications, and benefits. A total of 22 tools are categorised by their statistical family (parametric, non-parametric, and probabilistic) and level of analysis (transcript, exon, and event). The central challenges in quantifying alternative splicing include correct splice site identification and accurate isoform deconvolution of transcripts. Benchmarking studies show no consensus on tool performance, revealing considerable variability across different scenarios. Tools with high citation frequency and continued developer maintenance, such as DEXSeq and rMATS, are recommended for prospective researchers. To aid in tool selection, a guide schematic is proposed based on variations in data input and the required level of analysis. Additionally, advancements in long-read RNA sequencing are expected to drive the evolution of differential splicing tools, reducing the need for isoform deconvolution and prompting further innovation.

## Introduction

Alternative splicing (AS) can be best described as fine-tuning gene expression by rearranging exons and introns in pre-mRNA. With 90-95% of human multi-exon genes estimated to possess some form of alternative splicing, it is a widespread regulatory process in cellular biology (1). The cell utilises a large ribonucleoprotein (RBP) complex known as the spliceosome which is guided to target sites through the interaction of sequence elements (splice sites, enhancers & silencers and the polypyrimidine tract) and/or splicing factors. Pre-mRNA splicing can also occur without the splicesome as in the case of self-splicing group I & II introns, tRNA splicing and trans-splicing (2). This ultimately results in genome-wide transcript diversity and subsequently, measurable changes to protein functionality.

Previous research has uncovered the phenotypic consequences of alternative splicing in disease. In humans, clinical research has shown alternative splicing (AS) as a key instigator in several forms of cancer and neurodegenerative disorders (3–5). One notable discovery in Microtubule-associated protein tau's (MAPT) possession of mis-spliced isoforms causing abnormal TAU accumulation progressing to Alzheimer's disease (6). In cancer, numerous mis-spliced variants of tumour suppressors, apoptotic and angiogenic proteins have been discovered to contribute to tumour progression (7,8). Beyond clinical research, the utility of alternative transcripts for bioengineering purposes has been explored. For example, an alternatively spliced version of the transcription factor X-box binding protein 1 (XBP1) coexpressed in production cell lines has been shown to increase productivity in the biomanufacturing of recombinant proteins (9–11). In bio-agriculture, the CRISPR-mediated directed evolution of SF3B1 mutants (a spliceosomal component) in rice has improved crop traits through better resistance to splicing inhibitors (12). Increasingly, the value of AS in both clinical and biotechnology applications has been recognised; highlighting the need for robust bioinformatics pipelines to identify variants.

For prospective researchers to investigate AS, the transcriptomic data is usually generated using next-generation sequencing. Short-read RNAseq is the most commonly used experimental technique to interrogate a transcriptome owing to its versatility and cost-effectiveness (13,14). It involves sequencing short fragments of RNA molecules, providing insights into the respective expression levels of genomic features assembled from reference genomes. These features may be coding sequences, genes, transcripts, exons, introns, codons or even untranslated regions. A typical RNAseq pre-processing pipeline will consist of quality control (QC), read alignment & quantification before statistical analysis begins. QC assesses the quality of the raw fragmented reads using a standardised tool such as FastQC and trims low-quality reads or adaptor sequences (15). Then for alignment, a reference genome/transcriptome arranges the subsequent sequences into feature bins such as genes, transcripts, exons and coding sequences using software such as STAR or HISAT (16,17). Alignment files (usually in the form of Sequence Alignment Maps: SAMs) can then be quantified to these features using a quantification tool such as HTSeq, Salmon or featureCounts usually normalising for library size and sequencing depth (18–20). Depending on the purpose of analysis, normalisation may be scaled by total number of reads (CPM: Counts per Million), per length of transcript (TPM: Transcripts per Million), by paired-end fragments (RPKM: Fragments Per Kilobase of Transcript) or by using a median of ratios (DESeq2's method) (21). Commonly, a differential expression analysis will be performed at the gene or transcript level between groups of samples to identify statistically significant changes in expression. The pre-processing steps for RNA-seq have been extensively researched over many years, and there is a consensus within the community regarding the gold-standard set of tools. Projects like nf-core enable the execution of RNA-seq pre-processing pipelines with minimal intervention and limited bioinformatics expertise. (22). However, these tend to be focused on the use-case of conventional differential expression rather than the more bespoke AS pipelines as discussed here.

A growing repertoire of tools now annotate and quantify changes to splicing events. Quantification of features such as splice sites, and exon/intron junctions found in alignment files are commonly used to annotate splicing events. Although the true repertoire of splicing events is difficult to capture, conventional processes can be categorised into distinct groups. The most common events are exon skipping, retained introns, mutually exclusive exons, alternative 5' and 3' splice sites. More complex regulatory events involve genomic features beyond exons and introns, such as alternative promoter and polyadenylation sites, which result in varying mRNA 5' and 3' UTR ends. However, these events are seldom included in most bioinformatics analyses, tools such as CAGER (Cap Analysis of Gene Expression) and DaPars (Dynamic Analysis of Alternative PolyAdenylation from RNA-Seq) are available for niche research (23,24). Visualisation of AS is predicated upon the level of detail required in the analysis. If a highly detailed analysis of individual gene structure is needed, splice graphs, sashimi plots and junction maps are commonly used (25,26). To visualize changes to groups of transcripts, typically MA and Volcano plots are used much the same way as in differential expression level analysis (21).

## Main Text

## Current statistical methods for differential splicing

Commonly, researchers are interested in comparisons of two or more groups of samples known as differential analyses. Differential gene/transcript expression (DGE/DTE) of genes or transcripts involves taking raw read count data, normalizing or scaling it, and calculating whether the changes in expression levels between different biological groups are statistically significant. Differential transcript/exon usage (DTU/DEU), however, uses gene-level group modelling to assess whether the proportional use of the feature (exon or transcript) is statistically significant. Differential splicing events (DSE) on the other hand use a diverse array of statistical methods to quantify and infer splicing events. A comprehensive summary of differential splicing tools is described in the supplementary table (**Supplementary Table 1**) and in the following sections.

### Parametric & Mixed Methods

Differential expression analysis tools began in the early 2000s coinciding with the development of high throughput technologies such as microarrays. An early example was LIMMA (Linear Models for Microarray Data), developed by Gordon Smyth and colleagues in 2003, which utilises a linear regression framework and empirical Bayes techniques to identify differentially expressed genes (27). Whilst initially only utilised for microarrays, the functionality thus extended to RNASeq data and has been one of the most cited RNASeq methods. As the field shifted from microarray technology to RNASeq, methods were developed such as DESeq (Differential Expression Analysis for Sequence Count Data) and edgeR to capture the nature of count data better and improve modelling (21,27,28). A major change incorporated in DESeq2 was empirical Bayes-based shrinkage to improve gene-wise variance estimation enhancing accuracy **(Fig. 1)**. Secondly, GLMs (Generalized Linear Models) replaced the simple linear models as these were shown to adapt well to non-normally distributed count-based data (21). The flexibility of GLMs allowed algorithms to effectively deal with issues such as overdispersion, shrinkage, heteroscedasticity and covariates. To date, GLMs are usually fitted to the NB (Negative Binomial) distribution which confers some strong advantages. The NB distribution effectively captures overdispersion (the empirical variability of counts) and can handle a large excess of zero values commonly seen in transcript or exon-level count data. However, limma, DESeq2 and edgeR were not developed to specifically address the challenges of identifying AS.

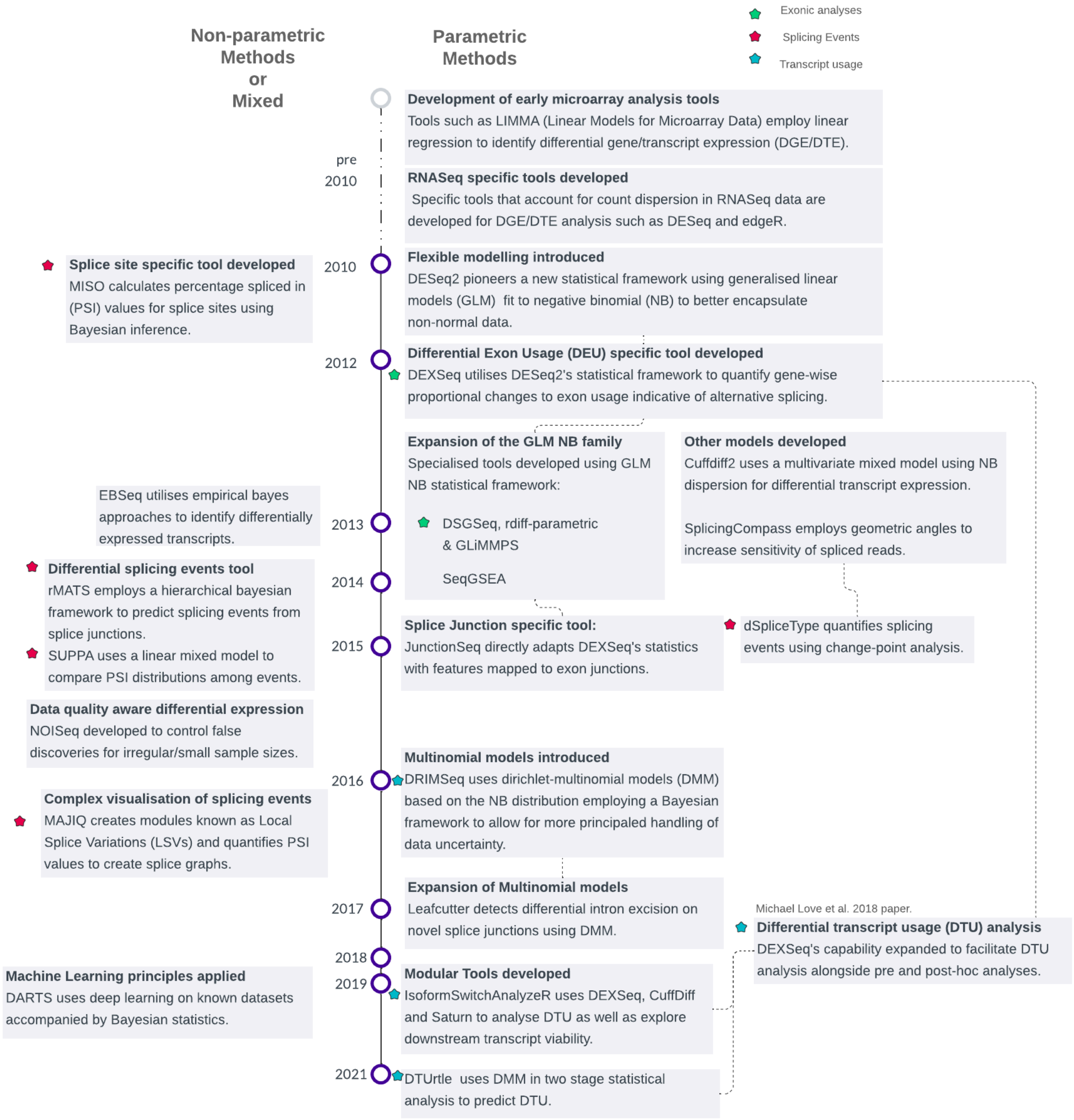

**Figure 1:** Timeline of statistical methods in differential splicing tool development. Here methods are split between parametric and non-parametric procedures and grouped into families. Parametric and non-parametric are roughly split based on the statistical procedures (namely for modelling or testing) in the supplementary table (**Supplementary Table 1**) however, a few methods may overlap and employ both.

In 2014, DEXSeq was introduced by Michael Love and colleagues, a framework based on DESeq2's GLM NB model becoming the de-facto tool for parametric splicing-based analysis. Instead of analysing gene-level differential expression, DEXSeq identifies exons within genes that exhibit significant changes in their usage across conditions. This is particularly useful for studying the exonic composition of alternatively spliced transcripts. The development of tools such as DSGseq, rDiff-parametric, JunctionSeq and SeqGSEA has expanded the functionality of the GLM NB family of differential splicing tools (29–32). DSGseq utilises a holistic approach considering splicing events not as individual elements but as comprehensive gene-wise splice graphs that more accurately reflect complex splicing dependencies (29). The tool rDiff-parametric on the other hand utilises isoform-specific loci such as restricted exonic regions to identify significant differences in isoform composition (30). The proposed advantage of this approach is in the smaller exonic regions rather than full isoform deconvolution. Assigning reads to isoforms is challenging because these transcripts are practically identical, making it difficult to definitively attribute a read from an overlapping region to a particular region without supplementary data. Therefore, full isoform deconvolution is significantly biased against genes with many isoform variants (33).

A few newer methods such as DRIMSeq and DTUrtle have progressed onto non-parametric or mixed Dirichlet Multinomial Models (DMM) which have been argued to capture better the complex variability of count data and better estimate isoform abundance (34,35) (**Fig. 1 & Supplementary Table 1**). Other methods such as IsoformSwitchAnalyzeR and some custom DEXSeq workflows now incorporate modularity allowing users a selection of bioinformatics tools for filtering, hypothesis testing and posterior calculations (36,37). An example of the usage of parametric analysis was in the discovery of a chimeric fusion transcript of PRKACA and DNAJB1 in a rare liver tumour FL-HCC (fibrolamellar hepatocellular carcinoma) using DEXSeq's differential exon usage framework (38). The discovery of differential exon usage of PRKACA's exons 2-10 and subsequent decreased usage of DNAJB1's exons 2-3 led the researchers to identify a chimeric transcript in FL-HCC patients. This demonstrated the utility of smaller exon-based analysis in identifying differences in transcript structure which would not be detected in larger gene or transcript-based analysis alone.

### Probabilistic & Non-parametric Methods

Non-parametric or probabilistic techniques such as MAJIQ, SUPPA, WHIPPET and rMATS frequently utilize Bayesian inference and/or probabilistic methodologies (26,39–41). By avoiding assumptions about the data's underlying distribution, these methods enable more sophisticated modelling. Consequently, in contrast to the predominantly standardized parametric exon/transcript-based techniques, event-based methods often showcase a broader array of statistical approaches **(Fig. 1)**. A few common features can be identified, however. Often the targets for event annotations are not labelled in gene-transfer format such as splice sites, exon/intron junctions and splicing quantitative trait loci (QTLs) which must be calculated. This then allows the "Percent spliced in" (PSI) to be calculated per exon, representing the ratio of the number of transcripts containing an alternative exon versus the total number of transcripts per any given splice site. By comparing PSI values, different splicing events can then be identified and explored through splice graphs and sashimi plots. An example of non-parametric tool usage was in the mapping of splicing events in the rice (Oryza sativa) transcriptome, revealing prevalent AS under deprived nutrient conditions (42). Importantly, this study utilised rMATs to reveal the underlying exon-intron structure of key nutrient transporter genes.

Some tools possess features for specific utility in certain scenarios. NOISeq is a non-parametric differential expression tool that is specifically designed to handle smaller numbers of biological replicates through its noise model (43). For more complex modelling, tools such as GLiMMPs (Generalized Linear Mixed Model for Pedigree Data with Population Substructure) employ mixed-effects models to account for both fixed and random effects such as genetic family substructure (44). Beyond splicing, the modular tool IsoformSwitchAnalyzeR facilitates analysis on spliced transcript quality such as Nonsense Mediated Decay (NMD) sensitivity, Intrinsically Disordered Regions (IDR) and protein domains (36). Increasingly, deep learning-based approaches are being utilised to improve the accuracy of differential splicing predictions leveraging publicly available RNASeq data such as with DARTs and Bisbee (45,46).

## Popularity & Developer Maintenance of Methods

To assess the academic popularity of tools, a citation and developer engagement analysis of original research articles within the Web of Science (WoS) domain and the respective GitHub website domains (if applicable). The assessment spanned from 2010 to 2024 and encompassed 19 original papers on various differential splicing tools. Notably, the citation counts for these splicing tools were considerably lower compared to conventional RNA-Seq differential expression analysis tools. For instance, while the general purpose DGE/DTE tool DESeq2 amassed a total of 35,887 citations during the same period, citations for differential splicing tools ranged from 7 to 1300 (**Fig. 2**). This discrepancy may pose challenges for researchers seeking resources and workflows specific to differential splicing analysis. Additionally, the importance of developer support cannot be understated, as it directly influences the usability and longevity of software tools. Notably, differential splicing tools such as DEXSeq, EBSeq, rMATS, SUPPA2, and MAJIQ (26,39,40,47,48) have shown increasing usage and ongoing developer engagement, as evidenced by their growing citation counts and sustained support (**Fig. 3; Fig. 4**). One possible explanation for the lower citation rates observed in exon/transcript-based methodologies could be the broader adoption of general-purpose differential expression workflows, like DESeq2 that can employ DTE (21). Researchers may prefer more explicit splicing event-based tools for targeted splicing analyses and defer to DTE for transcript-based analyses. While the nuances between DTU and DTE may not be a primary focus for many researchers, it is a distinction worth noting in the context of differential splicing analysis.

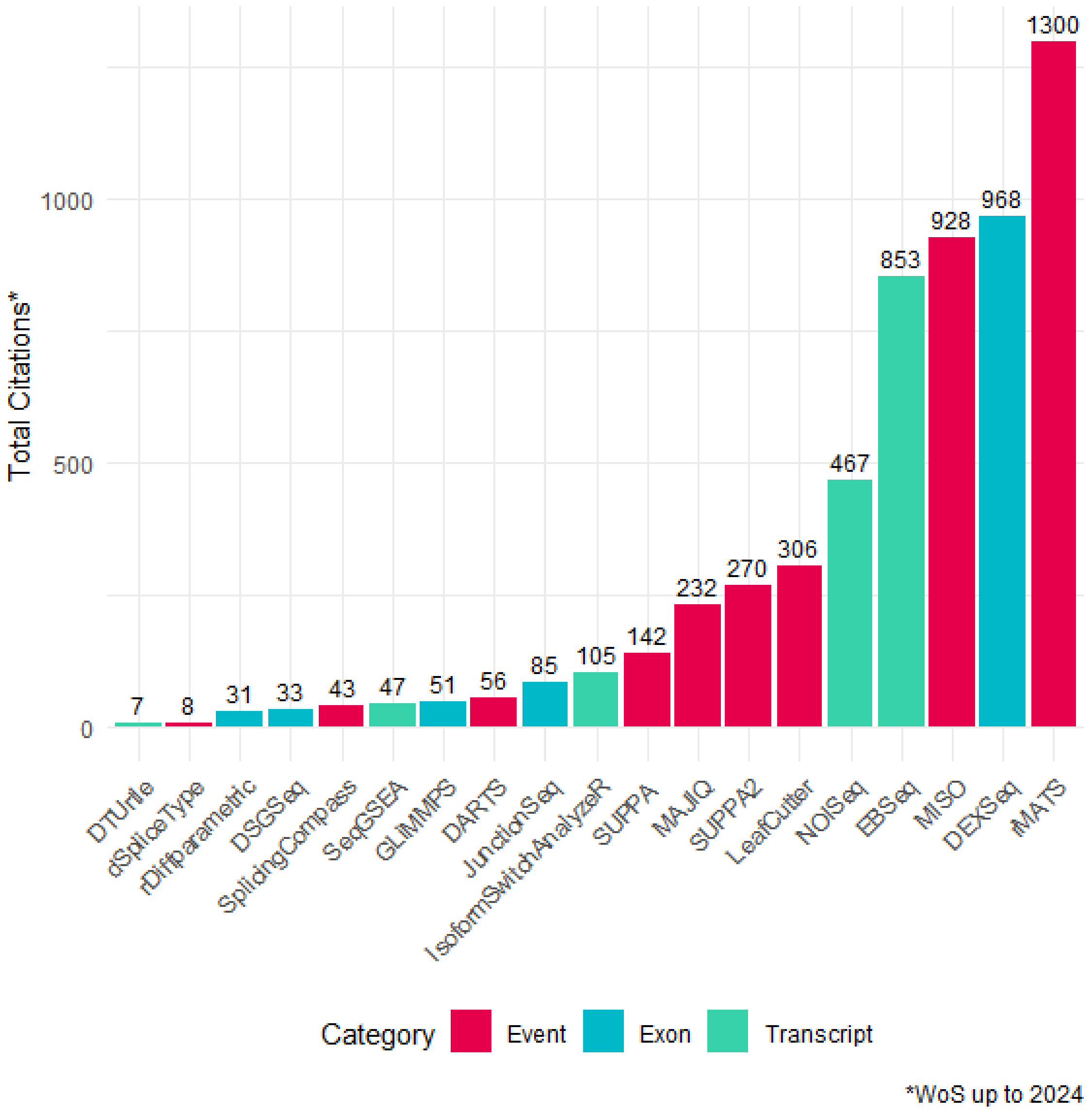

**Figure 2:** Total citation count for current differential splicing tools from the years 2010 up to 2024 from Web of Science (WoS). Only DRIMSeq's original paper was not found in the WoS platform and was thus excluded from citation frequency analysis. Certain data included herein are derived from Clarivate Web of Science.

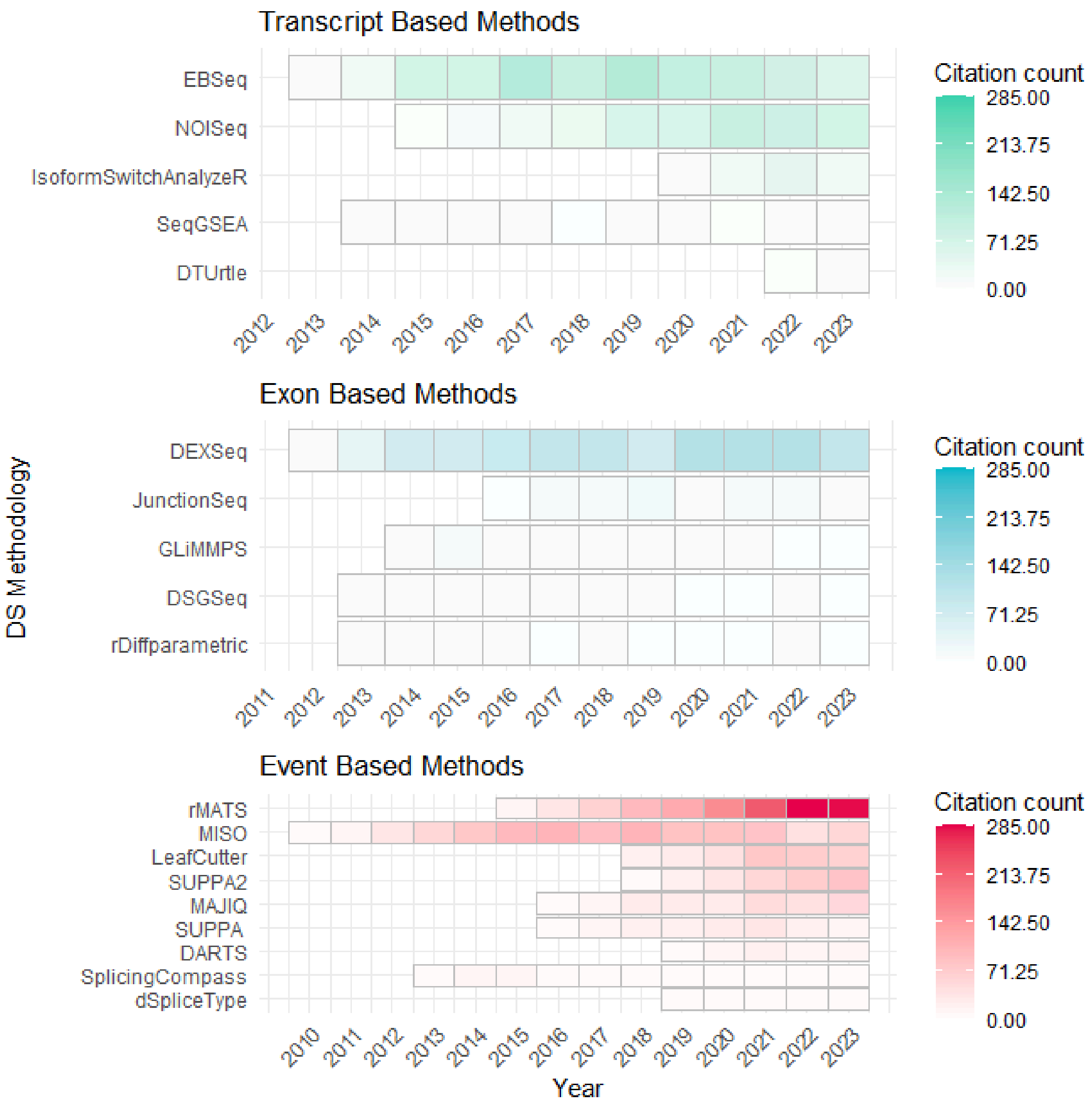

**Figure 3:** Citation frequency count for current differential splicing tools (excluding DRIMSeq) from the years 2010 up to 2024 from Web of Science (WoS). Certain data included herein are derived from Clarivate Web of Science. © Copyright Clarivate 2023. All rights reserved.

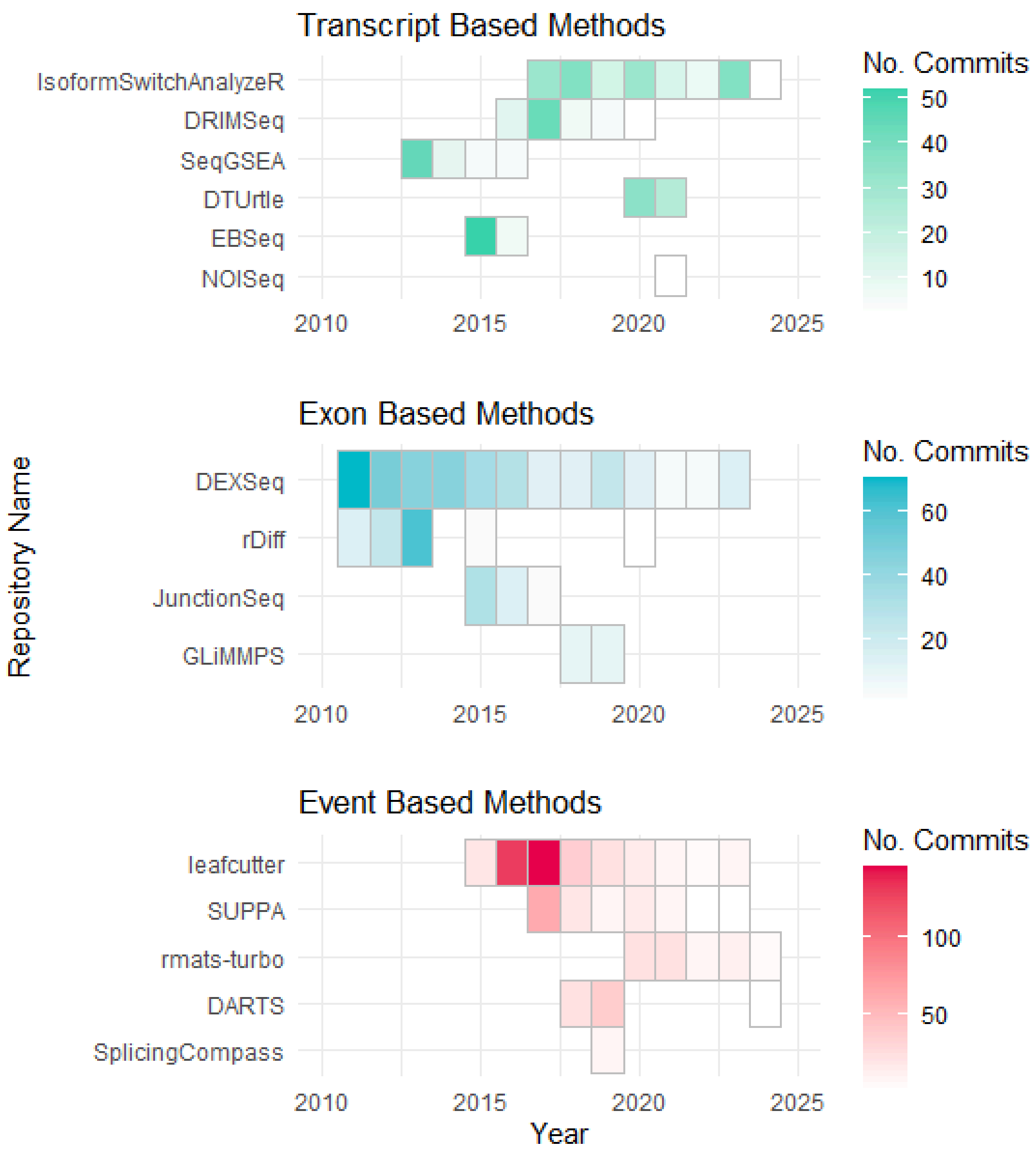

**Figure 4:** Commits to Github repositories from 2010 to 2024 split by category showing community-led maintenance by developers of tools. MAJIQ, MISO, DSGseq and dSpliceType do not have GitHub pages and were excluded.

The decision between exon/transcript-level (typically parametric) and event-level (typically non-parametric) analyses hinges on several factors, including the particular scientific inquiry, data accessibility, and the level of granularity required to address the research goal. In certain scenarios, integrating both methodologies could offer a more holistic understanding of splicing control mechanisms and their biological significance.

## Benchmarking of Methods is Difficult

To evaluate the quality of differential splicing bioinformatics tools, several benchmarks have been conducted to date. Benchmarking either the scientific accuracy or the computational power of methods can be challenging due to several factors. The main issue is the lack of ground truth to set as a reference to compare measurements to**.** Commonly, a small subset of experimentally validated splicing events is used as a gold standard to compare against. This was demonstrated in a recent systematic evaluation of 10 differential splicing tools in 2019, where a total of 62 qPCR-validated differentially spliced genes were tested (49). The results from this benchmark revealed weak consensus over tool quality as the performance was markedly different across the 4 human and mouse cancer datasets. This demonstrates another issue with these evaluations: inherent heterogeneity in RNASeq data. Often, the performance of methods will depend on the upstream RNASeq pre-processing steps such as in library size, sequence depth, positional bias and annotation quality. To mitigate these issues, some papers use simulated data to i) increase the number of differentially spliced genes to reference and ii)  achieve finer control over ground truth and variability within the data (50–52). One such benchmark used RSEM-based simulated data based on a human prostate cancer dataset (GSE22260: (53)) (51). Another comparison utilised a combination of experimental and simulated Arabidopsis heat shock RNASeq datasets using the Flux Simulator tool (54). However, it is important to note that simulated data lacks the complexity of typical biological data. Confounding factors such as outliers, and technical/procedural biases cannot be modelled in current simulations.

The consensus drawn from these three benchmarks is that the performance of differential splicing tools exhibits considerable variability depending on the outlined factors. The ongoing evolution and upkeep of tools by developers introduce a time-dependent aspect to benchmarking. Community-led maintenance efforts consistently enhance the functionality and reliability of tools over time. Rather than aiming for a singular optimal tool for differential splicing analysis, researchers should contemplate employing a suite of tools tailored to address specific inquiries.

## Method Recommendations

A diagram outlining optimal tool selection is provided to guide prospective alternative splicing (AS) researchers (**Fig. 5**). Initially, researchers should evaluate the scope and objectives of their analysis. For instance, if the aim is to identify known transcripts, it is advisable to opt for a parametric transcript-based tool like DEXSeq or DRIMSeq and execute a DTU study following Michael Love's protocol (37). Nonetheless, variations in experimental parameters such as sample size or covariate inclusion may necessitate alternative approaches.

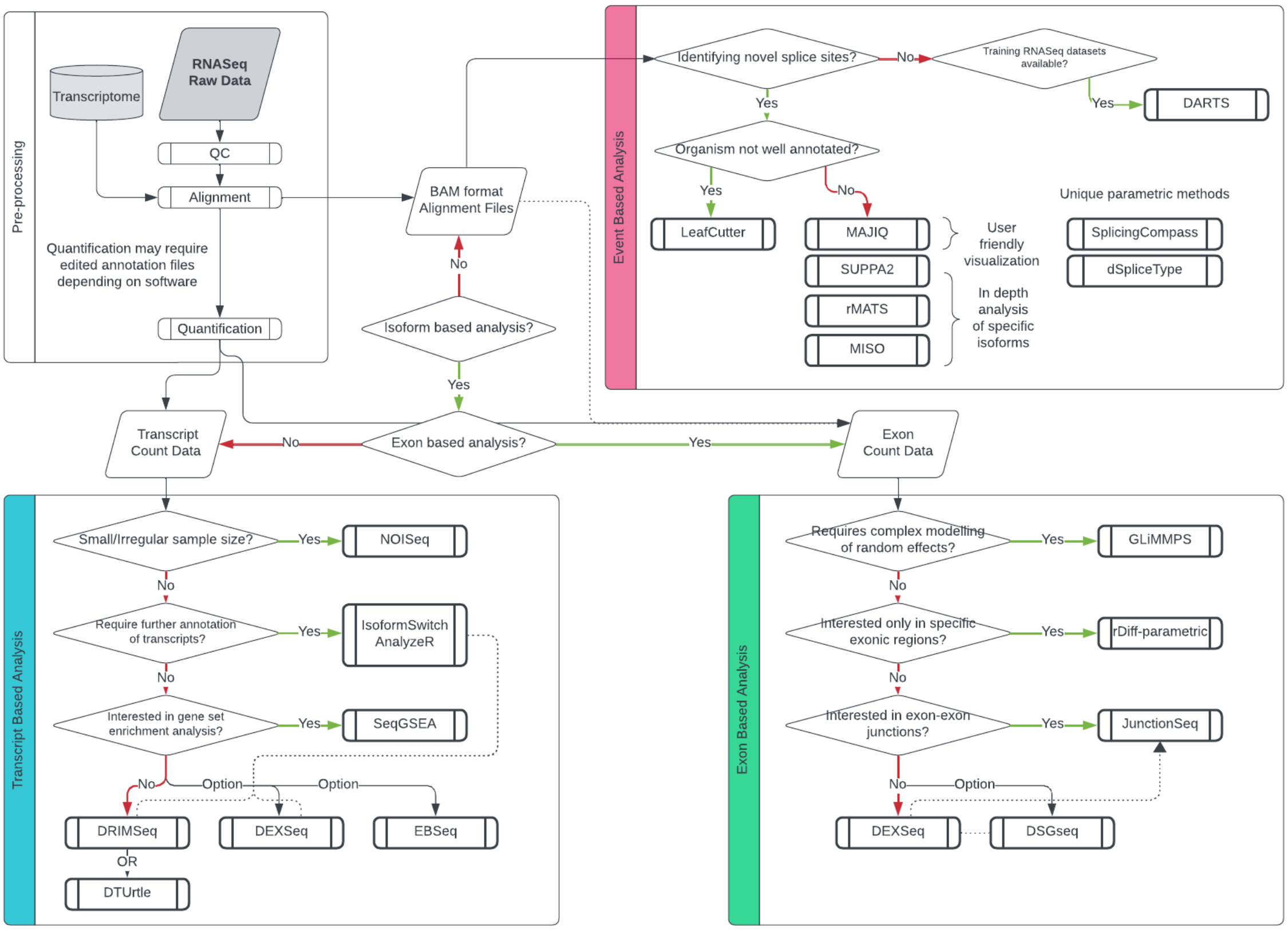

**Figure 5:** Decision tree of differential splicing for prospective researchers in the form of a decision tree in three branches representing the level of analysis. Transcript-based methods are in blue, exon-based methods are in pink and event-based methods are in yellow.

If the objective is to uncover novel transcripts, an exon-based parametric approach might be better suited. This choice circumvents the challenges associated with isoform deconvolution and the breadth of transcript annotation, given the smaller exonic regions. For general-purpose differential exon usage (DEU) analysis, DEXSeq remains the preferred protocol due to its robust and flexible statistical methods, as well as its actively maintained software (21). However, again intricacies within the data may prompt the usage of more specialised alternatives. Transcript and exon-based methods offer top-down visualizations such as MA/Volcano plots, heatmaps and proportional transcript/exon graphs. If the analysis aims to visualise the movement of exons/introns and splice sites, then an event-based protocol would be more appropriate. Generally, tools such as rMATs, SUPPA2 and MISO offer comprehensive and detailed splicing event analysis (39,40,55).

Commonly, sashimi plots are the best method to visualise splice junctions from aligned data with events annotated, although this can also be plotted separately in IGV (56). For user-friendly visualization, MAJIQ offers a summative HTML-based visualizer for complex events such as exitrons or orphan junctions (26). Another factor to consider is the annotation quality of the organism/tissue being studied. If researchers are not confident in the quality of annotations and would like annotation-free analysis, methods such as LeafCutter are a good alternative to conventional methods (57,58). Overall, event-based methods are more suited to advanced programmers owing to their use of command-line tools over interpreters that use IDEs (Integrated development environments). For most analyses, however, a DEU or DTU-based analysis is recommended for simple interpretability and robustness. Optional steps for AS-specific analyses can also be performed to enhance the data quality. For example, Portcullis enables the accurate filtering of false splice junctions that are often incorrectly characterized by common aligners (59).

## Discussion

While the repertoire of tools to accommodate differential splicing analysis has grown in the past two decades, they are ultimately limited by the capabilities of the RNASeq technology available to date. Since 2010 however, the development of nanopore sequencing technology such as Oxford Nanopore Technologies (ONT)  and PacBio's single-molecule real-time (SMRT) has facilitated the development of long-read RNAseq  (60–62). Long read lengths typically fall within the range of 10kb to 100kb, with ultra-long read lengths now up to 1-2 Mb (63). The main benefit this technology confers is the ability to bypass the aforementioned deconvolution issue stemming from multiple mapping and reconstruct full-length transcript isoforms in a single read. This can not only more accurately identify known transcripts but also novel or splice variants as well as fusion genes. Most current parametric DS tools can therefore be utilised in long-read-based analyses. A recent study utilized IsoformSwitchAnalyzeR's DEXSeq-based DTU workflow on ONT long reads, demonstrating the capability of current long-standing methods on long-read data (21,36,64). This was facilitated through long-read custom annotation of the transcriptome using TALON to identify novel transcripts (65). Additionally, specific novel technologies such as LIQA have been developed to analyse long reads (66).

Long-read RNAseq still possess notable disadvantages, however. Early on, long-read RNASeq possessed error rates of 10-20% (67,68). The development of HiFi sequencing by PacBio using circular consensus sequencing has since reduced the error rate to a reported 0.5% (69). While the development of deep-learning algorithms such as DeepConsensus has sought to push HiFi accuracy further bringing it on par to short read (70). However, this is still highly dependent on the depth of sequencing. The most efficient error correction method involves hybridising the analysis with short-read RNAseq methods (71). This ultimately means that while accuracy can now be brought to close to 99.5%, error correction drives the cost of long-read RNAseq methods up significantly. The field is progressing towards optimal error correction and is now focusing on lowering costs which is currently the largest hurdle for practical use for common research.

As interest in alternative splicing grows, researchers have access to an expanding array of tools. Advances in statistical methods and longer RNA sequencing read lengths are overcoming technical limitations. This leads to more precise transcript alignment and reduces the need for complex computational steps. With workflows becoming streamlined and modular, platforms like Nextflow enable researchers to create tailored pipelines for their specific goals and data types (72). These developments promise a brighter future for alternative splicing analysis, facilitating a deeper exploration of transcriptomic regulation and its functional significance.

**Acknowledgements:**

This work was supported by funding from *Lonza Biologics Inc.*

https://www.lonza.com/

I extend my gratitude for their financial assistance in facilitating this research project. The funders had no role in study design, data collection and analysis, decision to publish, or preparation of the manuscript.

**Data Availability Statement:**

All data used to create figures 2 and 3 were retrieved from the Web of Science (WoS) data repository using the original papers cited in the references,  GitHub repositories were cloned on 20.02.2024 for the creation of Figure 4. All R code used to generate the figures are available on request.

The GitHub repositories are listed here.

https://github.com/TobiTekath/DTUrtle
https://github.com/areyesq89/DEXSeq
https://github.com/lengning/EBSeq
https://github.com/Xinglab/rmats-turbo
https://github.com/yarden/MISO
https://github.com/davidaknowles/leafcutter
https://github.com/comprna/SUPPA
https://github.com/Xinglab/DARTS
https://github.com/ConesaLab/NOISeq
https://github.com/kvittingseerup/IsoformSwitchAnalyzeR
https://github.com/Wanvdphelys/SeqGSEA
https://github.com/hartleys/JunctionSeq
https://github.com/Xinglab/GLiMMPS
https://github.com/gosianow/DRIMSeq
https://github.com/ratschlab/rDiff
https://github.com/KoenigLabNM/SplicingCompass
https://github.com/thelovelab/DESeq2

## Supporting Information

| Date of Publication | Tool Name | Specificity | Level of analysis | Statistical Summary | Statistic Type |
|---|---|---|---|---|---|
| 2010 | MISO (55) | Splicing | Event | Uses a probabilistic Bayesian framework to model and detect splicing events. Percentage spliced-in (PSI) values were calculated for splice sites or alternative exons. Hypothesis testing uses posterior probabilities and Bayes factors. | Probabalistic |
| 2010 | DESeq2 / DESeq (21) | General | Transcript /Gene | Uses a generalized linear model (GLM) fit to the negative binomial (NB) distribution. Addresses gene-specific dispersion variation. Uses a variance stabilizing transformation (VST) to address Poisson-like distribution. Hypothesis testing through the Wald test is used between conditions to test for differential expression. False discovery rate (FDR) control using the Benjamini-Hochberg procedure. | Parametric |
| 2012 | DEXSeq (47) | Splicing | Exon/Transcript | Adapted from DESeq2 and edgeR's generalized linear model (GLM). Models read counts per exon using a GLM fitted to the negative binomial (NB) distribution. Addresses dispersion variation through shrinkage. Uses a variance stabilizing transformation (VST) to address Poisson-like distribution. Hypothesis testing through likelihood-ratio test (LRT) to test for differential usage. False discovery rate (FDR) control using the Benjamini-Hochberg (BH) procedure. | Parametric |
| 2012 | Cuffdiff2 (73) | General | Transcript | Multivariate mixed model fit to a beta negative binomial (NB) distribution with multi-read correction. Trimmed mean of m-values (TMM) normalization applied. Hypothesis testing using the likelihood ratio test (LRT) to test for differential expression. False discovery rate control using the Benjamini-Hochberg (BH) procedure. | Parametric |
| 2013 | Splicing Compass (74) | Splicing | Event | Employs geometric angles to increase sensitivity on exon read counts. Designed for the detection of more complex splicing events. Hypothesis testing using a one-sided t-test and Benjamini Hochberg (BH) correction to control false | Parametric |
| | | | | discovery rate (FDR). | |
| 2013 | DSGseq (29) | Splicing | Exon | Uses a negative binomial (NB) model on exon counts to detect differential spliced genes. Hypothesis testing using NB statistics without classical P value significance. | Parametric |
| 2013 | GLiMMPs (44) | Splicing | Exon | Uses a generalized linear mixed model (GLMM) to fit the negative binomial (NB) distribution to account for phylogenetic relationships as a source of random effects. hypothesis testing using a likelihood ratio test (LRT). detects splicing quantitative trait loci (QTLs). | Parametric |
| 2013 | riff-parametric (30) | Splicing | Exon (Regions) | Uses a negative binomial (NB) model on smaller exonic regions that are indicative of relative isoform abundance. Hypothesis testing was performed using diff.parametric equation per exonic region with Bonferroni corrections. Has an option for rDiff-nonparametric if there is no gene annotation. | Mixed |
| 2013 | EBSeq (48) | Splicing | Transcript /Gene | Uses hierarchical modelling under an empirical Bayesian framework under a negative binomial distribution (NB). Hypothesis testing using posterior probabilities to evaluate probabilities per gene. | Probabilistic |
| 2014 | rMATs (40) | Splicing | Event | Detects splicing events using splice site junctions. Uses a joint probability model to compare splicing events between conditions using both the normal and negative binomial distribution. Employs a hierarchical Bayesian framework to estimate percent-spliced-in (PSI) values for each splicing event. Hypothesis testing is conducted using a likelihood ratio test (LRT). False discovery rate (FDR) control is applied using benjamini-hochberg (BH) correction. | Non-parametric |
| 2014 | SeqGSEA (32) | Splicing | Transcript | Adapted from DESeq2 and DSGSeq's generalized linear model (GLM) on negative binomial (NB) distributions. Incorporates a ranked-based strategy for overrepresented gene sets. Employs gene set enrichment analysis (GSEA). | Parametric |
| 2015 | SUPPA (39) | Splicing | Event | Detects splicing events using event annotations. Uses a linear mixed model (LMM) under the beta distribution. Percent spliced-in (PSI) values are calculated per splicing event annotated. Hypothesis testing uses the Wilcoxin-Mann-Whitney U test to compare PSI distributions. False discovery rate (FDR) control is applied using Benjamini-Hochberg (BH) correction. Applies the Pareto principle. | Non-parametric |
| 2015 | dSpliceType (75) | Splicing | Event | Utilises a multivariate statistical model for splicing events. Change-point analysis followed by a parametric statistical test using the Schwarz information criterion (SIC). | Parametric |
| 2015 | JunctionSeq (31) | Splicing | Exon/Splice Junctions | Adapted from DEXseq's generalized linear model (GLM) fitted to the negative binomial (NB) distribution. Read counts are instead mapped to splice/exon-exon junctions. Uses Bayesian estimation to compute posterior probabilities for splicing changes. False discovery rate (FDR) control using the Benjamini-Hochberg (BH) procedure. | Parametric |
| 2015 | NOISeq (43) | Splicing | Transcript/Gene | Specifically designed for small or irregular sample sizes. Uses a non-parametric model under a Bayesian approach. Hypothesis testing using posterior probabilities to evaluate probabilities per gene. | Non-parametric |
| 2016 | MAJIQ (26) | Splicing | Event | Detects local splicing variations (LSVs) encompassing event complexity. Uses beta distribution modelling for inclusion levels under a Bayesian framework. Utilises posterior distributions for multiple hypothesis testing. Percentage spliced index (PSI) calculated using likelihood estimates. | Non-parametric |
| 2016 | DRIMSeq (34) | Splicing | Transcript | Uses a Dirichlet-multinomial model (Bayesian framework) fitted to a negative binomial distribution. Constructs a "splicing index," which represents the relative abundance of alternative splicing isoforms. An exact test is used to test for differential expression. False discovery rate (FDR) control using the Benjamini-Hochberg procedure. | Parametric |
| 2017 | LeafCutter (57) | Splicing | Event | Detects splicing quantitative trait loci (SQTL) using splice junctions with percentage spliced in (PSI) values. | Parametric |
| | | | | Differential intron excision. Dirichlet-multinomial generalized linear model. Hypothesis testing using permutation testing under a Bayesian framework. | |
| 2018 | WHIPPET(41) | Splicing | Event | Uses entropy based statistics. Employs a Bayesian framework to estimate the posterior probability of different splicing outcomes based on observed RNA-Seq read counts, allowing for the identification and quantification of splicing variations. | Non-parametric |
| 2019 | DARTS (46) | Splicing | Event | Uses a trained deep neural network (DNN) model accompanied by empirical Bayesian statistics for hypothesis testing. | Non-parametric |
| 2019 | IsoformSwitchAnalyzeR (36) | Splicing | Transcript | Modular package. Primarily utilises DEXSeq model to test for isoform switches but includes Cuffdiff or Saturn options. Facilitates further annotation of features such as open reading frames (ORF), coding potential and nonsense-mediated decay sensitivity of transcripts. | Parametric |
| 2021 | DTUrtle (35) | Splicing | Transcript | Adapted directly from DRIMSeq's Dirichlet-multinomial model (DMM). Hypothesis testing uses a likelihood ratio test (LRT). Two-stage statistical procedures were applied using stage's false discovery rate (FDR) corrections through the Benjamini-Hochberg (BH) procedure. | Parametric |

**Supplementary Table 1:** Statistical summary of current differential splicing tools with categorisation into parametric, non-parametric or probabilistic based on the testing statistic employed. The summary captures the model and distribution type, hypothesis testing methods, and corrective procedures. Categorised are the three types of splicing analysis by output: exon-based, transcript-based and event-based.